\documentclass[12pt]{article}

\usepackage{epsfig}
\usepackage{cite}
\usepackage{amsmath, amssymb, amsfonts}
\usepackage{color}
\usepackage{latexsym}
\usepackage{graphicx}

\def\be{\begin{equation}}
\def\ee{\end{equation}}
\def\ba{\begin{eqnarray}}
\def\ea{\end{eqnarray}}

\def\bdm{\begin{displaymath}}
\def\edm{\end{displaymath}}
\def\la{~\mbox{\raisebox{-.6ex}{$\stackrel{<}{\sim}$}}~}
\def\ga{~\mbox{\raisebox{-.6ex}{$\stackrel{>}{\sim}$}}~}
\def\bq{\begin{quote}}
\def\eq{\end{quote}}

 at 10truept

\newcommand{\p}{\partial}
\newcommand{\de}{\partial}

\renewcommand{\[}{\left[}
\renewcommand{\]}{\right]}
\renewcommand{\(}{\left(}
\renewcommand{\)}{\right)}
\newcommand{\vk}{\vec{k}}

\newcommand{\eps}{\epsilon}

\newcommand{\Mpl}{m_{\mathrm{Pl}}}

\newcommand{\fnl}{f_{\mathrm{NL}}}

\newcommand{\bea}{\begin{eqnarray}}
\newcommand{\eea}{\end{eqnarray}}

\newcommand{\bi}{\begin{itemize}}
\newcommand{\ei}{\end{itemize}}

\newcommand{\order}{{\cal O}}
\newcommand{\beq}{\begin{equation}}
\newcommand{\eeq}{\end{equation}}
\newcommand{\beqa}{\begin{eqnarray}}
\newcommand{\eeqa}{\end{eqnarray}}
\newcommand{\mpl}{\Mpl}

\def\la{~\mbox{\raisebox{-.6ex}{$\stackrel{<}{\sim}$}}~}
\def\ga{~\mbox{\raisebox{-.6ex}{$\stackrel{>}{\sim}$}}~}

\newcommand{\calR}{\mathcal{R}}
\newcommand{\vx}{\vec{x}}
\newcommand{\Z}{\mathbb{Z}}
\newcommand{\cK}{{\cal K}}

\def\ltap{\ \raise.3ex\hbox{$<$\kern-.75em\lower1ex\hbox{$\sim$}}\ }
\def\gtap{\ \raise.3ex\hbox{$>$\kern-.75em\lower1ex\hbox{$\sim$}}\ }
\def\gl{\ \raise.5ex\hbox{$>$}\kern-.8em\lower.5ex\hbox{$<$}\ }
\def\roughly#1{\raise.3ex\hbox{$#1$\kern-.75em\lower1ex\hbox{$\sim$}}}

\begin{document}

\thispagestyle{empty}
\begin{flushright}
September 2017 \\
BRX-TH 6323\\
CERN-TH-2017-181
\end{flushright}
\vspace*{.5cm}
\begin{center}

{\Large \bf Monodromy inflation at strong coupling: $4\pi$ in the sky }\\

\vspace*{.75cm} {\large Guido D'Amico$^{a, }$\footnote{\tt
damico.guido@gmail.com}, Nemanja Kaloper$^{b, }$\footnote{\tt
kaloper@physics.ucdavis.edu}, and Albion Lawrence$^{c, }$\footnote{\tt
albion@brandeis.edu}}\\
\vspace{.3cm} {\em $^a$Theoretical Physics Department,
CERN, Geneva, Switzerland}\\
\vspace{.3cm}
{\em $^b$Department of Physics, University of
California, Davis, CA 95616, USA}\\
\vspace{.3cm} {\em $^c$Martin Fisher School of Physics, Brandeis University, Waltham, MA 02453, USA}\\

\vspace{1cm} ABSTRACT
\end{center}
We present a simple effective field theory formulation of a general family of single field flux monodromy models for which strong coupling effects at large field values can flatten the potential and activate operators with higher powers of derivatives. These models are radiatively and non-perturbatively stable and can easily sustain $\ga 60$ efolds of inflation. The dynamics combines features of both large field chaotic inflation and $k$-inflation, both of which can suppress the tensor amplitude. Reducing the tensor-scalar ratio below the observational bound $r \lesssim 0.1$  while keeping the scalar spectral index $n_s$ within experimental bounds either yields equilateral nongaussianity $f_{NL}^{eq} \simeq {\cal O}(1)$, close to the current observational bounds, or ultimately gives very small $r$.

\vfill \setcounter{page}{0} \setcounter{footnote}{0}
\newpage

Models of inflation with superplanckian inflaton ranges can be remarkably simple and predictive \cite{Linde:1983gd,Freese:1990rb}.  They give the largest possible primordial tensor fluctuations, potentially detectable by CMB polarization experiments.  However, control over the relevant $\geq \mpl$ field range requires additional model-building
input. A promising control mechanism is monodromy \cite{Silverstein:2008sg,McAllister:2008hb,Kaloper:2008fb,Kaloper:2011jz}. In recent work, two of us have proposed that generic monodromy inflation models can be given a simple field theory formulation based on the dual massive 4-form gauge theory \cite{Kaloper:2016fbr} (see also \cite{Marchesano:2014mla,Hebecker:2015zss}). The gauge symmetries -- a compact $U(1)$ symmetry for the 4-form and a discrete phase space gauge symmetry for the dual scalar -- protect slow roll inflation from both perturbative and nonperturbative corrections.

The gauge field mass $m$ is a dimensionful coupling which controls the theory below the cutoff $M$. It appears in the combination $m\phi$, with $\phi$ the inflaton field, so we take $\frac{m\phi}{M^2}$ as a dimensionless coupling up to factors of 2 and $\pi$. A weakly coupled theory produces a canonical kinetic term and quadratic potential \cite{Kaloper:2008fb,Kaloper:2011jz,Kaloper:2016fbr,Kaloper:2014zba}, predicting tensor to scalar ratio $r \sim 0.16$. The current upper bound $r \lesssim 0.1$ on primordial tensors thus rules out weak coupling.  In this letter we discuss the strong coupling regime of the EFT \cite{Kaloper:2016fbr}, using Na\"ive Dimensional Analysis (NDA) \cite{Manohar:1983md,Gavela:2016bzc} to write down the most general gauge-invariant, ghost-free EFT for energies below the cutoff scale $M$. Factors of $4\pi$ open up an interval of strong coupling below the cutoff $M$, with irrelevant operators comparable to marginal and relevant couplings,  while slow roll remains protected from UV physics.

Strong coupling provides at least two significant effects: nontrivial ``flattened" potentials as seen in string theory constructions \cite{Silverstein:2008sg,McAllister:2008hb,Dong:2010in,McAllister:2014mpa}, and higher-derivative terms such as those in k-inflation \cite{ArmendarizPicon:1999rj,Garriga:1999vw}.  We identify two dynamically and observationally distinct slow-roll inflationary phases, depending on which class of effect dominates. In one phase, the higher-derivative terms are dynamically suppressed, and inflation is the standard potential-driven slow-roll. Either flattening provides a reduction of tensor power, while increasing the spectral index $n_s$; or the potential plateaus out \cite{Dubovsky:2011tu,Nomura:2017ehb}, in which case $n_s$ remains close to the observed value, but $r$ is at most at the edge of detectability and the field range is at most ${\cal O}(\mpl)$.  In the other phase, the higher derivative terms are important, suppress the tensor power and enhance ``equilateral" nongaussianity to $f_{NL}^{eq} \simeq {\cal O}(1)$, close to the current observational bounds. In this class, bounds on $f_{NL}^{eq}$ imply a {\it lower}\ limit on $r$. 

We will address the details regarding which phase dominates in a given model, and how a transition from one to another can occur in future work. Here we will study the phases separately and explore the parameter ranges consistent with observations and naturalness.

The low energy EFT of the flux monodromy inflation model of \cite{Kaloper:2016fbr} is the most general massive abelian three-form gauge theory, including all the irrelevant operators allowed by gauge symmetry.  The action is\footnote{Using the equations of motion to eliminate terms with $\p^k F$ factors.}
\ba \label{corrections}
{\cal L}^{\rm (full)}  &=&  - \frac{1}{48} F_{\mu\nu\lambda\sigma}^2 - \frac{m^2}{12} (A _{\mu\nu\lambda} - h_{\mu\nu\lambda})^2   - \sum_{n>2} \frac{a'_n}{M^{2n-4}}\tilde F^{n}  \nonumber\\
& & - \sum_{n > 1} \frac{a''_n }{M^{4n-4}} m^{2n} (A_{\mu\nu\lambda} -h_{\mu\nu\lambda})^{2n}   - \sum_{k\ge1, \, l\ge 1} \frac{a'''_{k , l}}{M^{4k+2l-4}} m^{2k}
(A_{\mu\nu\lambda}-h_{\mu\nu\lambda})^{2k} \tilde F^{l}  \, ,	\ea
where $A$ is the gauge field 3-form, $F=dA$, ${\tilde F} = {}^*F$, $b$ a two-form St\"uckelberg gauge field with field strength $h=db$, and $m$ plays the role of both the gauge field mass and the Stueckelberg mode coupling.
By gauge symmetry and the Goldstone Boson Equivalence Theorem (GBET), any power of $A$ not covered by a derivative must be multiplied by the same power of $m$ \cite{Kaloper:2016fbr}. Finally, $M$ is the cutoff of the low energy EFT.

For the purposes of analyzing inflationary dynamics, we will work in the dual frame in which the longitudinal mode of $F$ is replaced by a compact scalar: schematically, $F \sim \epsilon (m \phi + Q), mA \sim \epsilon \partial \phi$ \cite{Kaloper:2016fbr}.  Here $Q = N q$, where $q$ is the fundamental 4-form charge; $N \in \Z$; and $\phi \equiv \phi + f$ where $mf = q$ \cite{Kaloper:2011jz}.  We will use the NDA framework to provide a more refined description of the  dimensionless coefficients in (\ref{corrections}):
there are crucial factors of $1/4\pi$ ensuring that the EFT including loop corrections below the cutoff $M$ is natural \cite{Manohar:1983md,Gavela:2016bzc}.  Such factors will give us a substantial inflaton range at strong coupling. The rules are:
\begin{itemize}
\item Replace $\phi$ by the dimensionless quantity $\frac{4\pi \phi}{M}$;
\item Include the overall normalization $\frac{M^4}{(4\pi)^2}$ to normalize the Lagrangian;
\item Include factorials in the denominators of (\ref{corrections}) to account for the symmetry factors in the physical S-matrix elements.
\end{itemize}
The upshot is that the strong coupling scale is $M_s = M/\sqrt{4\pi}$, above which irrelevant operators can be activated without also activating the UV degrees of freedom. The resulting effective Lagrangian, valid to the scale $M$, is\footnote{Again, we remove $\p^k\phi$ terms via the equations of motion.}
\ba \label{correctionsscalarnorm}
{\cal L} &=&  - \frac{1}{2} (\partial_\mu \phi)^2 - \frac{1}{2} (m \phi + Q)^2  - \sum_{n>2} c'_n \frac{(m\phi + Q)^{n}}{ n! (\frac{M^2}{4\pi})^{n-2}} \, \nonumber \\
&& -  \sum_{n>1} c''_n \frac{(\partial_\mu \phi)^{2n}  }{2^n n!(\frac{M^2}{4\pi})^{2n-2}} - \sum_{k\ge 1, \, l \ge 1}  c'''_{k,l}\frac{(m\phi + Q)^{l} }{{2^k k! l! (\frac{M^2}{4\pi})^{2k+l-2}}}
(\partial_\mu \phi)^{2k}  \, .
	\ea
where $c_n \sim {\cal O}(1)$ by naturalness of the strong coupling expansion\footnote{The $c_n$'s depend on  $m\phi + Q$ through the RG logs. At large $n$ they may grow as $\sim n!$
by proliferation of graphs with fixed external lines, such
 that (\ref{correctionsscalarnorm}) is not an analytic function for all values of $\phi, X$. 
}, unless they are suppressed (i.e., zero) by additional symmetries. Next, define the inflaton as $\varphi = \phi + Q/m$. Since $Q = 2\pi  N q$ with $N \in Z$, the inflaton {\it vev} $\varphi$ can vary over a range $\gg \mpl$, while {\it all fundamental parameters including $f$ are below $m_{pl}$}.
The action (\ref{correctionsscalarnorm})\ can then be formally written as
\be \label{kinflation}
{\cal L} =  K\Bigl(\varphi, X \Bigr) - V_{eff}(\varphi)  =  \frac{M^4}{16\pi^2} {\cal K}\Bigl(\frac{4\pi m \varphi}{M^2}, \frac{16\pi^2  X}{M^4} \Bigr) - \frac{M^4}{16\pi^2} {\cal V}_{eff}(\frac{4\pi m \varphi}{M^2}) \, ,
\ee
where $X \equiv - (\partial_\mu \varphi)^2$, and ${\cal K}, {\cal V}_{eff}$ should be understood as 
asymptotic series, well approximated by finitely many terms. This is precisely the form of the action of the k-inflation model of \cite{ArmendarizPicon:1999rj,Garriga:1999vw},
with variables $\varphi, X$ normalized by $\frac{4\pi m}{M^2}$ and $\frac{16\pi^2}{M^4}$ respectively, 
and kinetic and potential functions given by (truncated) asymptotic series, with natural coefficients.
It is the natural EFT for single-field monodromy inflation.

As we have stated, Eq. (\ref{kinflation}) supports two distinct phases.  In the first, the terms nonlinear in $X$ are dynamically suppressed, while the third term can sum to a flattened potential supporting standard potential-driven slow roll.  The effective action for $\varphi$ is
\be
{\cal L} = -\frac12 {\cal Z}_{eff}(\frac{4\pi m\varphi}{M^2}) (\partial_\mu \varphi)^2 - \frac{M^4}{16 \pi^2}{\cal V}_{eff}(\frac{4\pi m\varphi}{M^2}) \label{eq:twod}\ ,
\ee
where ${\cal Z}_{eff}$ and ${\cal V}_{eff}$ are dimensionless functions, and the normalizations of the arguments reflect NDA and gauge symmetries. When $m\varphi < M^2/4\pi$, we are in the weak coupling regime: ${\cal V}_{eff}$ is a power series in $\frac{4\pi m\varphi}{M^2}$, and reduces to a quadratic potential with small corrections discussed in \cite{Kaloper:2008fb,Kaloper:2008qs,Kaloper:2014zba}. The kinetic corrections are also small, as seen in (\ref{correctionsscalarnorm}), reproducing the classic scenario  in \cite{Linde:1983gd}. If the last 60 efolds of inflation occur in this regime, the computed tensor-scalar ratio $r \simeq 0.16$ is in conflict with cosmological observations \cite{Ade:2013ydc}.

Observations thus force our model to reside at strong coupling $M^2/4\pi < m\varphi$ during the epoch at which the observed CMB fluctuations are generated. The effective theory remains valid at least to $m\varphi < M^2$, leaving a window in which strongly coupled inflation can occur without exciting UV degrees of freedom. In this regime, the potential can leave the quadratic regime and flatten out \cite{Silverstein:2008sg,McAllister:2008hb,Dong:2010in,McAllister:2014mpa}.  The precise form of flattening will be determined by the UV completion, which fixes the form of ${\cal Z}_{eff}$ and ${\cal V}_{eff}$. Note that the slow roll parameters
\be
	\eps= \frac{m_{pl}^2}{2} \left(\frac{V_{eff}'}{V_{eff}}\right)^2 \sim \left(\frac{4\pi m m_{pl}}{M^2}\right)^2 \, , ~~~~~~~~~~	\eta = m_{pl}\big|\frac{V_{eff}''}{V_{eff}}\big| \sim \left(\frac{4\pi m m_{pl}}{M^2}\right)^2 \, , \label{eq:slowrollcond}
\ee
are small so long as $\frac{4\pi m m_{pl}}{M^2} \ll 1$: slow roll is compatible with strong coupling for super-Planckian excursions of $\varphi$. The standard slow roll equations of motion give $16\pi^2 X/M^4 \sim \order(\eps)$, so that this phase is self-consistent: we can approximate $\cK$ by the linear term.

One variant of flattening is induced by wave function renormalization of the inflaton. Canonically normalizing the field $\chi = \int d \varphi {\cal Z}^{1/2}_{eff}(\frac{4\pi m \varphi}{M^2})$, and this often
yields the effective potential of the form $V_{eff} \sim \chi^p$, $p< 2$. Such a theory can easily accommodate a sufficient number of efolds. When the strong coupling regime can be approximated by power law, then $N_{\rm total} \sim \int d\chi \frac{V_{eff}}{\mpl^2 \partial_\chi V_{eff} } \sim (\chi /\mpl)^2$. If $M^2/m > \chi > M^2/(4\pi m)$,  this regime supports $N_{\rm strong} \la (M^2/m \mpl)^2 \simeq (4\pi)^2 N_{weak}$ efolds, where $N_{weak} \la \frac{1}{64 \pi^2}\frac{M^4}{ m^2 \mpl^2}$ is the number of efolds in weak coupling.
So long as $N_{weak} \ga 1$, $N_{\rm strong} \ga (4\pi)^2 \sim 150$.  Requiring $N_{weak} < 50$ implies
$m^2 > \frac{M^4}{3200 \pi^2 \mpl^2}$. Demanding slow roll (\ref{eq:slowrollcond})\ in the strong coupling regime gives $\partial_\chi^2 V_{eff} < V_{eff}/3 m_{pl}$, which yields:
\be
  \frac{M^4}{3200 \pi^2 \mpl^2} < m^2 <  \frac{M^4}{48 \pi^2 \mpl^2}  \, .
  \label{massconst}
\ee
During the evolution inflation proceeds from strong coupling to weak coupling, and the exact duration
of the weak coupling regime is controlled by $m$, but as long as $m$ obeys (\ref{massconst}) it is shorter than
50 efolds. This guarantees that the observable CMB anisotropies were generated during the strong coupling regime, such that $r \lesssim 0.1$. 
However, since the spectral index for the potential $V_{eff} \sim \chi^p$ is $n_s = 1 - \frac{p+2}{2{\cal N}}$ at ${\cal N}$ efolds before the end of inflation, in this regime
$p$ cannot be arbitrarily small in order to fit the observations of $n_s \simeq 0.96$ for $50 \lesssim {\cal N} \lesssim 60$.

A more extreme reduction of $r$ can occur if
in the approximation of the potential ${\cal V}_{eff}$ there are cancellations which cause
the potential at large $\phi$ to reach a plateau. This may happen in strongly coupled regimes with many particles \cite{Dubovsky:2011tu}. In this example, one finds an effective potential such as
${\cal V}_{eff} \sim (1- \frac{1}{1+ (4\pi c m \varphi/M^2)^p})$. Fitting the observations implies $\frac{M^2}{4\pi c m} \lesssim {\cal O}(1) \mpl$ \cite{Nomura:2017ehb}, and as a result after fitting $n_s$ to $\sim 0.96$ the tensor to scalar ratio is small, $r \sim {\rm few} \times 10^{-4}$ \cite{Dubovsky:2011tu}. The field variation during inflation is also small, $\delta \varphi \lesssim {\cal O}(1) \mpl$. Such models appear perfectly consistent with observations, although their values of $r$ might be too small to be observable by the next generation of polarization experiments.\footnote{Note that in these models the curvature of the potential increases as we approach the end of inflation, and while our EFT is valid during inflation, reheating could involve the UV degrees of freedom. We leave this issue for future work.}${}^{,}{}$\footnote{Another possibility, discussed in \cite{Creminelli:2014nqa}, is a potential of the form $A + B \varphi^p$ with $A,B$ positive.  This allows a wider range of values of $r,n_s$ consistent with observation, but requires additional degrees of freedom to discharge $A$ and reheat, so we leave this possibility aside.}

A second phase, also yielding a reduction of $r$, occurs when higher derivative operators become important, yielding a variant of k-inflation.
The cosmological observables in k-inflation have been studied extensively in \cite{Garriga:1999vw,Gruzinov:2004jx,Chen:2006nt}. The higher derivatives modify both the linear and nonlinear perturbations \cite{Garriga:1999vw}.
From the terms in the action quadratic in fluctuations, one finds that the speed of sound of the scalar perturbations $c_s$ is different from unity, so that  the tensor-scalar ratio is $r = 16 c_s \eps$, where $\eps$ is the usual slow-roll parameter~\footnote{For small $c_s$, it may be important to consider corrections logarithmic in $c_s$, as discussed in \cite{Baumann:2014cja}.}.
The inflationary consistency relation is $r = - 8 c_s n_T$, where $n_T$ is the tensor spectral index.
In essence, the higher derivatives improve the slow roll in the scalar field equation, without changing the gravitational field equations; this is reflected in the modification of the consistency condition.
Meanwhile, terms in the action that are cubic in fluctuations \cite{Gruzinov:2004jx,Chen:2006nt} can lead to non-negligible equilateral nongaussianities.
We will present the results of the numerical investigation of monodromy k-inflation, to illustrate the tension between small $r$, a spectral index consistent with observations, and small nongaussianities in this framework. This tradeoff is the main observational implication of monodromy in this regime.

We will take inflation be potential-driven, as required by the combination of slow roll and the null energy condition \cite{dkalong}. We consider the case that the potential has a wide smooth regime of variation towards the minimum, instead of a sharply ending plateau. Furthermore, we focus on the regime in which higher derivative terms dominate over the quadratic ones both in controlling the background and the perturbations. Finally,  for illustrative purposes, we consider the specific kinetic energy function $K = {\cal Z} X + \tilde {\cal Z} \frac{16\pi^2 X^2}{M^4}$.

In strong coupling, the zero modes satisfy $16\pi^2 X/M^4 \gtrsim 1$ and $16 \pi^2 V_{eff}/M^4 \gtrsim 1$. The background field equations are
\be
3 \mpl^2 H^2 = V_{eff} \, , ~~~~~~~~~~~~~  6H \dot \varphi \tilde {\cal Z} X = - 4\pi M^2 m V_{eff}' \, .
\label{background}
\ee
There are a number of possible histories depending on how the derivative terms evolve relative to the potential.  We will explore this in detail in \cite{dkalong}, and summarize the results here.
If the derivatives decay faster than the potential, higher-derivative inflation turns over to inflation with quadratic kinetic terms after some number of efolds controlled by $m$. If the derivatives decay more slowly than the potential or grow, higher derivative inflation can continue on all the way to the boundary of strong coupling.  At this point $16\pi^2 X/M >1$ but $16\pi^2 V_{eff}/M^4 = 1$ and the slow roll conditions will be at least briefly violated. If $m$ is too large, after the transition to weak coupling the potential will be unable to support any slow roll, and inflation will end. For smaller $m$, after the transition to weak coupling the dynamics will dissipate the derivatives and slow roll inflation with a quadratic potential can restart. Inflation therefore can have at least two stages: strongly coupled slow roll with large derivatives and a short weakly coupled stage with small derivatives (whose duration may be so short as to be negligible, depending on $m$). The weakly coupled regime must be shorter than 50 efolds in order to allow strong coupling effects to flatten the potential and reduce $r$ to meet the observational bounds, implying $m$ still must satisfy (\ref{massconst}). Happily, suppression of $r$ by higher derivatives can occur without increasing $n_s$ relative to weakly coupled inflation, which allows a theory consistent with observations. However
nonlinearities induced by higher derivatives yield nongaussianities: this in turn again yields a {\it lower} bound on $r$ in this regime.

We consider the effective theory for the inflaton as in \cite{Cheung:2007st}, neglecting the mixing with gravity. This is a good approximation for energies $E \gg \sqrt{\eps} H$ and $c_s^2 \gg \eps$, and yields an almost scale-invariant spectrum of fluctuations with small nongaussianities, as observations require. In this limit, working in the gauge where the spatial metric is unperturbed and flat, the field 
perturbation $\pi$ is defined by $\varphi(t, \vx) = \varphi_0(t + \pi(t, \vx))$;  the spatial
curvature perturbation is ${\cal R} = - H \pi$. Expanding Eq.~\eqref{kinflation} up to third order, and keeping only the terms lowest in derivatives, we get
\be \label{Lpi}
S = - \int dt d^3\vec x \, a^3 \Mpl^2 \dot{H} \[ \frac{1}{c_s^2} \dot{\pi}^2 - \frac{(\de_i \pi)^2}{a^2}
+ \(\frac{1}{c_s^2} -1 \) \( \dot{\pi}^3 + \frac{2}{3} c_3 \dot{\pi}^3
- \dot{\pi} \frac{(\de_i \pi)^2}{a^2} \)
\]  \, .
\ee
In terms of the EFT of monodromy \eqref{kinflation}, the speed of sound is $c_s^2 = \partial_X {\cal K} / (\partial_X  {\cal K} + 2 X \partial_X^2  {\cal K})$, and $c_3(1/c_s^{2}-1) = 2 X^2 \partial_X^3  {\cal K}/ \partial_X  {\cal K}$. 
When $X > \frac{M^4}{16\pi^2}$, this means $c_s^2 \sim \frac{M^4}{32\pi^2 X} \frac{{\cal K}''}{{\cal K}'}$ and $c_3(1/c_s^{2}-1) = \frac{512 \pi^4 X^2}{M^8} \frac{{\cal K}'''}{{\cal K}'}$, or $c_3 = \frac{16\pi^2 X}{M^4} \frac{{\cal K}'''}{{\cal K}''}$.
Whether ${\cal K}$\ and its derivatives are ${\cal O}(1)$, large, or small when $y = 16\pi^2 X/M^4 > 1$ depends on the UV theory governing the large-$y$ asymptotics.

The perturbation amplitude leading to CMB fluctuations is set at horizon crossing $\lambda^{-1} \simeq H$, and the scale at which the background breaks time translation symmetry is $f_\pi \propto \sqrt{\dot \varphi}$. In the strong coupling regime with large derivative terms, $f_{\pi}^4 = 2 \Mpl^2 \dot{H} c_s$, and (using Eq. (\ref{background})) $\Mpl^2 H^2 \simeq \frac{M^4}{48\pi^2}$.
The gaussian scalar power spectrum is $\Delta_{\calR}^2 \propto (H/f_{\pi})^2$.  $2 \rightarrow 2$ scattering of perturbations shows that unitarity of (\ref{Lpi})\ is saturated at 
$\Lambda^4(t) \simeq 16 \pi^2 \Mpl^2 \dot{H} c_s^5/(1-c_s^2)$. When
$X > M^4/16\pi^2$ in strong coupling, $c_s^2 \lesssim 1$. 
In this regime the higher derivative terms are important both for perturbations and for the background  evolution. 

Note that our class of monodromy EFTs coupled to gravity can remain valid even when the Goldstone theory (\ref{Lpi}) breaks down. This could happen when $\epsilon/c_s^2$ is large, or when the cutoff $\Lambda(t)$ of the action (\ref{Lpi}) drops below the fundamental 
cutoff $M$ and even below $f_\pi$ for small $c_s$. The former requires the inclusion of graviton modes; the latter requires that (\ref{Lpi}) should be UV-completed by our EFTs for energies $\Lambda < E < M$. 
Nonetheless, (\ref{Lpi}) remains valid for $c_s$ down to $\sim \epsilon^{1/2}$, and for energy scales $\omega < \Lambda(t)$. As long as  $H/\Lambda(t) \simeq  [\frac{3}{16\pi^2 \epsilon c_s^5} (H/\mpl)^2 ]^{1/4} < 1$, (\ref{Lpi}) can be invoked to describe the horizon crossing of perturbations, allowing (\ref{Lpi}) to $c_s \ga {\rm few} \times 10^{-2}$. A more detailed  treatment is required to fully describe the generation of fluctuations \cite{dkalong}. The concerns about strong coupling for (\ref{Lpi}) when $\Lambda > f_\pi$ \cite{Baumann:2014cja} 
are addressed for us by
the full covariant EFT of monodromy which is valid all the way to the real cutoff $M$, and $f_\pi < M$. The theory {\it is} strongly coupled, but it is still under control below the cutoff $M$, as we explained. 

The leading nongaussianities come from the 3-point function. There are two operators in (\ref{Lpi}) which source them, leading to three-point functions of the form
\be
\langle \Phi_{\vk_1} \Phi_{\vk_2} \Phi_{\vk_3} \rangle
= (2 \pi)^3 \delta_D(\vk_1+\vk_2+\vk_3)
\frac{6 \Delta_{\Phi}^2}{(k_1+k_2+k_3)^3} \[ f_{\rm NL}^{(1)} F_1(k_1, k_2, k_3) + f_{\rm NL}^{(2)} F_2(k_1, k_2, k_3) \] \, ,
\ee
where
\be
f_{NL}^{(1)} = - \frac{85}{324} \(\frac{1}{c_s^2}-1 \) \, , \qquad
f_{NL}^{(2)} = - \frac{10}{243}
\(1-c_s^2 \) \(\frac{3}{2} + c_3 \) \, .
\ee
Here $F_1$ and $F_2$ are induced by $\dot{\pi} (\de \pi)^2$ and $\dot{\pi}^3$, respectively \cite{Chen:2006nt,Senatore:2009gt,Ade:2013ydc}.

In our simple example $c_3 = 0$. More generally, while $c_3$ will be of order $1/c_s^2$, unless ${\cal K}'''/{\cal K}'' \gg {\cal K}'/{\cal K}''$ the numerical prefactors in $f^{(1)}_{NL}$\ are larger than those in $f^{(2)}_{NL}$\ by a factor of 5. A lower bound on $c_s$ restricts the range of cubic nongaussianities to a narrow strip in the $f^{(1)}_{NL}, f^{(2)}_{NL}$ plane \cite{DAmico:2014ptm}, similarly to the models \cite{Alishahiha:2004eh,DAmico:2012khf,DAmico:2012ji}.

\begin{figure}
	\centering
	\includegraphics[width=8cm]{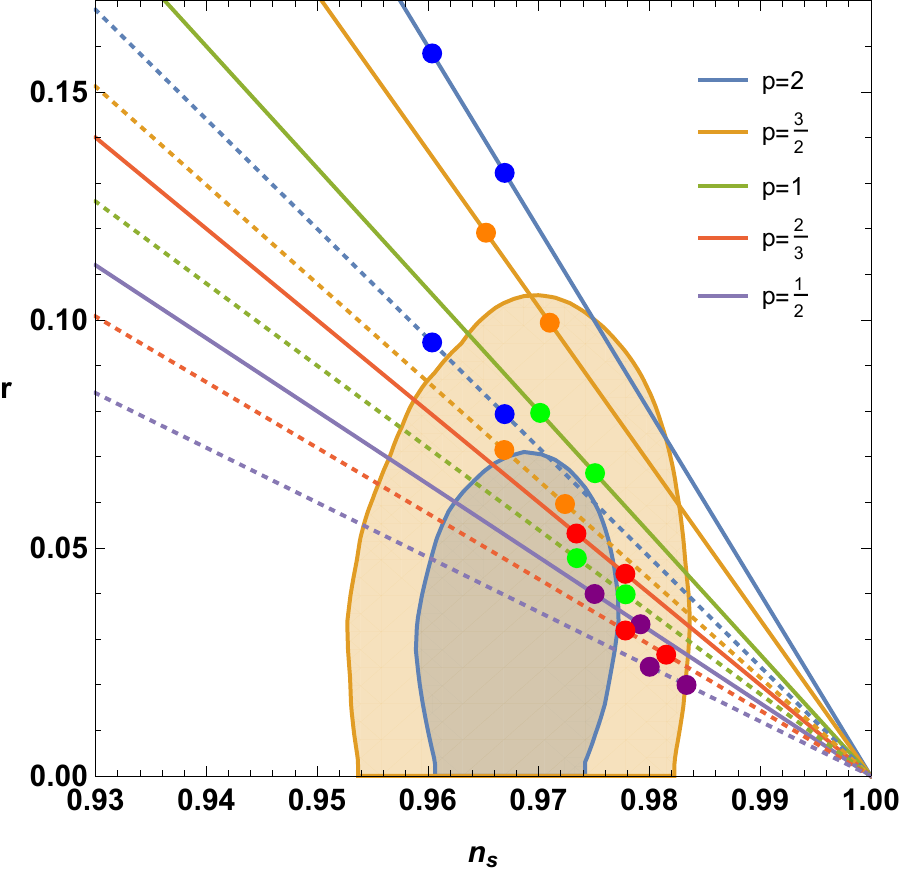}
	\caption{Planck constraints in the $n_s-r$ plane for $\Lambda$CDM model for power-law potentials $V_{eff} \sim \phi^p$, with $p$'s listed in the legend, and the kinetic term quartic in fields. The solid lines are for $c_s = 1$, and the dashed lines for $c_s = 0.6$. The colored circles denote normalizations at 50 (left) and 60 efolds (right), respectively.}
	\label{fig:rvsns}
\end{figure}

\begin{figure}
	\centering
	\includegraphics[width=8cm]{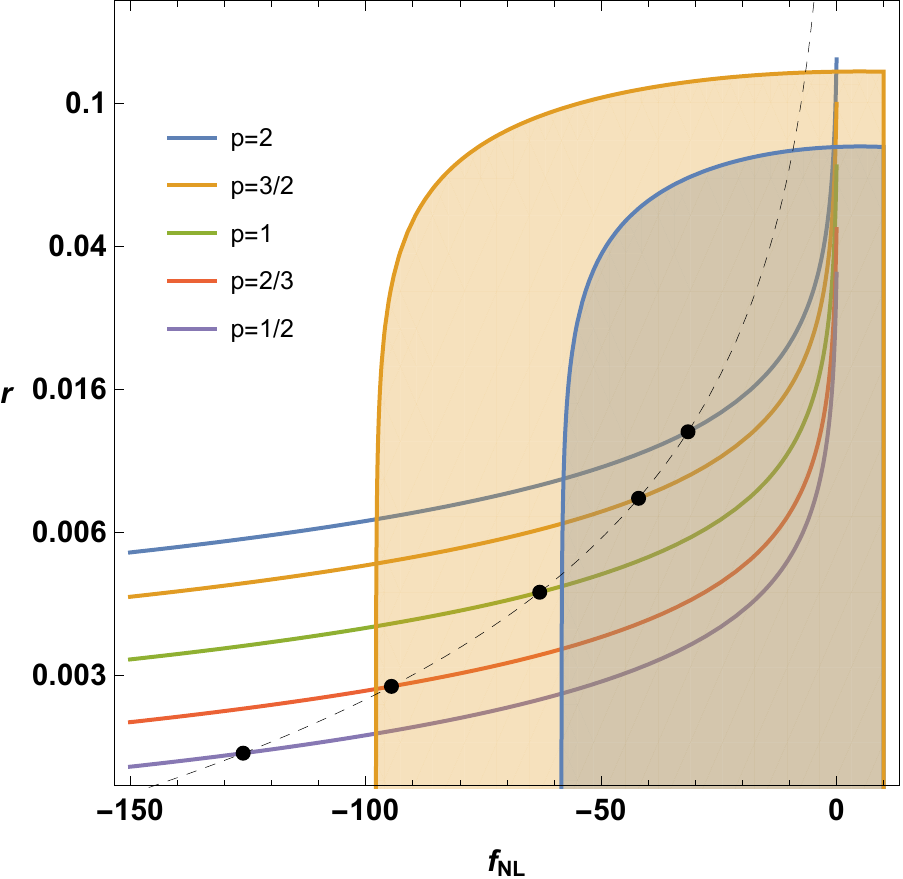}
	\caption{Tensor-to-scalar ratio $r$ v.s. equilateral nongaussianity $\fnl$ for inflation with a potential $V_{eff} \sim \phi^p$, with fixed $p$ as indicated in the legend; $c_s$ varies along each curve from unity at the top right to small values on the left. The kinetic term is quartic in fields. Dashed line corresponds to $c_s^2 \sim \eps$, and above it the Goldstone action (\ref{Lpi}) is not a reliable description of perturbations (a more complicated description may nevertheless exist).}
	\label{fig:rvsfnl}
\end{figure}

Current observations may already cut into the space of our strongly coupled EFTs. If we use $c_s^2 = \partial_X {\cal K} / (\partial_X  {\cal K} + 2 X \partial_X^2  {\cal K})$, we could attain $c_s$ as low as 
$1/\sqrt{32 \pi^2} \simeq 0.056$, when $X \la M^4$, while the observational bound
$f_{NL}^{(1)} \la 50$ implies $c_s \ga 0.07$ \cite{Ade:2013ydc}. However for so small $c_s$ the perturbative description of fluctuations using (\ref{Lpi}) may be unreliable; yet the full theory may be valid, since $X \la M^4$. Thus a more complete understanding of the perturbations in this regime is needed, and our full theory provides a good starting point for developing it. In Figures (\ref{fig:rvsns}) and (\ref{fig:rvsfnl}) we present the predictions for the observables $r$, $n_s$ and $\fnl$, for generic power-law potentials in the regimes where (\ref{Lpi}) holds.
We stress that the precise form of the potential and higher derivative operators depends on the details of the UV completion. From a bottom-up point of view, we would use data to determine whether our EFT is realized, and given this class of theories, to identify the dominant operators during inflation.

In summary, we have shown that strong coupling dynamics of flux monodromy based on massive $3$-form gauge theories naturally leads to flattened potentials at large field values, which are perturbatively and nonperturbatively stable. A key feature of the dynamics is that the curvature of the potential remains small even for large field values. This follows from
gauge symmetries of the theory that restrict the potential to the form given in (\ref{kinflation}), where the mass parameter $m$ is stable since it is also the gauge coupling. For sufficiently low $m$, the slow roll conditions can be realized naturally, evading the problems of modular inflation \cite{Banks:1995dp}. In the case (\ref{eq:twod}), flattening consistent with largeish $r$ such as given by power law potentials
either tends to raise the spectral index $n_s$ above current bounds, or it gives a plateau potential that meets all observational bounds but predicts a very small $r$.

However, our EFT has a phase which combines both strongly coupled large field chaotic inflation and $k$-inflation, in which the total reduction of the primordial tensor amplitude arises both due to flattening and to higher derivatives. In turn, large derivative contributions increase nongaussianities
and so in this case satisfying the observational bound $r \lesssim 0.1$ on tensors while retaining $n_s$ near $0.96$ requires equilateral nongaussianity $f_{NL}^{eq} \simeq {\cal O}(1)$, close to the current observational bounds. This still leaves a broad range of possible values for $r$ that can be probed by future observations. This makes our natural monodromy models an excellent benchmark for future observations.

\section*{Acknowledgements}

We would like to thank Zackaria Chacko, Roni Harnik, Matt Kleban, Eva Silverstein and Alexander Westphal for useful conversations on this and related subjects.  N.K. thanks the CERN Theory Division for hospitality in the course of this work. A.L. thanks Nordita and the ICTS for hospitality during the course of this work.
N.K. is supported in part by DOE Grant DE-SC0009999. A.L. is supported in part by DOE grant DE-SC0009987.

\bibliography{oddflationrefs.bib}

\end{document}